\documentclass[onecolumn,pra,aps,preprintnumbers,amsmath,amssymb]{revtex4}
\usepackage{units}

\begin{document}

\markboth{H. Wunderlich and M. B. Plenio}
{Estimating purity and entropy in graph state experiments}

\title{ESTIMATING PURITY AND ENTROPY IN STABILIZER STATE EXPERIMENTS}

\author{Harald Wunderlich$^{1,2,3}$}
\email{h.wunderlich@physik.uni-siegen.de}
\author{Martin B. Plenio$^{1,2}$}
\email{m.plenio@imperial.ac.uk}
\affiliation{$^1$Institute for Mathematical Sciences, Imperial College London,
53 Prince's Gate, SW7 2PG London, UK \\
$^2$QOLS, Blackett Laboratory, Imperial College London, Prince Consort
Road, SW7 2BW London, UK \\
$^3$Fachbereich Physik, Universit\"at Siegen, Walter-Flex Str. 3,
57068 Siegen, Germany}

\begin{abstract}
Many experiments in quantum information aim at creating graph states. Quantifying
the purity of an experimentally achieved graph state could in principle be
accomplished using full-state tomography. This method requires a number of
measurement settings growing exponentially with the number of constituents
involved. Thus, full-state tomography becomes experimentally infeasible even
for a moderate number of qubits.
\\
In this paper we present a method to estimate the purity of experimentally achieved graph states with simple measurements. The observables we consider
are the stabilizers of the underlying graph. Then, we formulate the problem
as: "What is the state with the least purity that is compatible with the measurement
data?" We solve this problem analytically and compare the obtained bounds
with results from full-state tomography for simulated data. 
\end{abstract}

\keywords{Graph state; Stabilizer; Purity.}

\maketitle

\section{Introduction}    
Stabilizer states \cite{1} (notably including cluster states \cite{1a}) 
represent a major class of entangled states. They form the resource for 
a number of applications in quantum information, such as quantum 
computing\cite{2}, quantum error correction\cite{3}, and quantum cryptography\cite{4}.
Because of the usefulness of cluster states, a considerable effort has been
devoted to their theoretical studies as well as to their experimental implementation.
To date, photonic cluster states of four\cite{5} and six qubits\cite{6} have been experimentally
demonstrated, and their realization with trapped ions is actively pursued\cite{7}.

The experimental progress made in the area of cluster states effects the need
for sophisticated methods for the estimation of the system's properties.
A natural problem which arises is the fact that the density matrix grows
exponentially with the number of qubits involved. So far, it has already
been shown that fidelities and entanglement in cluster state experiments can be estimated using a number of observables linear in the number of qubits\cite{8,9}.
The observables of choice are the so-called stabilizers (see e.g. Ref. \cite{10}
for an introduction to the stabilizer formalism), which
are given by
\begin{equation}
K_j = X_j Z_{N_j},
\end{equation}  
where $X$ denotes the usual Pauli x operator acting on qubit $j$, and $Z_{N_j}
$ denotes the Pauli z operator acting on neighbours of $j$ defined by the
underlying graph. Note that for a graph of $n$ qubits, the set $\{K_1,\dots,K_n\}$
generates an abelian group, called the stabilizer group. Graph states form
the simultaneous eigenvector with eigenvalue $+1$ to these stabilizers. 

In this paper we will show how to estimate the purity $tr(\rho^2)$ in cluster
state experiments. We will consider the generators of the stabilizer group
as the observables. Similar to the estimation of entanglement in Refs. \cite{11}
(cfr. Ref. \cite{13} for a review of other entanglement estimation
methods), the problem is formulated as: "What is the quantum state
with the lowest purity, and which is compatible with the measurement data?" Mathematically, this question is the following quadratic optimization problem:
\begin{align}
\label{eq:Pmin}
P_{min} = min_{\rho} \left[ tr(\rho^2) : tr(K_i \rho) = a_i, \rho \ge 0 \right].
\end{align}
The paper is structured as follows: in Sec. \ref{sec:Symmetries} we will utilize
the symmetries allowed by the observables to restrict the optimization to
stabilizer diagonal states.  Then, we provide an
exact analytical solution to the above optimization problem in Sec. \ref{sec:Lowest-Purity}.
The following section proceeds with a discussion of the quality of the obtained
solution by comparing it with results from full-state tomography from simulated
data of a noisy system. Sec. \ref{sec:entropy} discusses the estimation of
the von-Neumann entropy from stabilizer measurements, and provides
an analytical lower bound on the entropy.
 
\section{Estimating purities}
\subsection{Symmetries of the stabilizers}
\label{sec:Symmetries}
As mentioned in the introduction, the natural observables in cluster state
experiments are given by the stabilizers. Let us assume the goal of an experiment
was the creation of a pure cluster state, with stabilizers $K_j$, $j \in \{1,\dots,n\}$.
We denote the measurement outcomes by $a_j = tr(\rho K_j)$. Note that these
expectation values are invariant under rotations of the stabilizer group.
By twirling over the stabilizer, one may therefore restrict to states of
the form
\begin{align}
\label{eq:rho}
\rho = \frac{1}{2^n} \sum_{i_1,\dots,\i_n=0}^1 c_{i_1 \dots i_n} K_1^{i_1}
\dots K_n^{i_n}.
\end{align} 
Here, the following twirling protocol was utilized:
\begin{align}
\rho' \longrightarrow \rho = \frac{1}{2^n} \sum_{i_1,\dots,i_n=0}^1 K_1^{i_1} \dots K_n^{i_n} \rho' K_1^{i_1} \dots K_n^{i_n}.
\end{align}
The coefficients in the stabilizer decomposition of $\rho$ are subject to
the normalization constraint $c_{0 \dots 0}=1$ and measurement outcomes $a_j
= tr(\rho K_j) = c_{\tilde{j}}$, where $\tilde{j}$ denotes a bit string of
zeros, but $1$ in position $j$. Furthermore, any valid density matrix must
be positive-semidefinite. The stabilizer decomposition of $\rho$ provides
a convenient way to calculate the eigenvalues of $\rho$. Since the stabilizers
mutually commute, and the spectrum of each term in the decomposition is simply
given by $\{+1,-1\}$, one finds the following expression for the eigenvalues:
\begin{align}
\label{eq:lambda}
\lambda_{j_1 \dots j_n} = \frac{1}{2^n} \sum_{i_1,\dots,i_n=0}^1 (-1)^{\sum_l i_l\cdot
j_l}c_{i_1 \dots i_n}.
\end{align} 

\subsection{Lowest purity compatible with measurement data}
\label{sec:Lowest-Purity}
In this section we determine the lowest purity compatible with measurement
data in cluster state experiments. So we calculate the exact solution to the optimization problem (\ref{eq:Pmin}). As seen in the previous section, it
is legitimate to restrict to stabilizer diagonal states of the form (\ref{eq:rho}).
Then, the problem boils down to solving the following positive quadratic program:
\begin{align}
minimize & \quad & P(c) = \frac{1}{2^n} \sum_{i_1,\dots,\i_n=0}^1 c_{i_1
\dots i_n}^2
\end{align}
\begin{align}
s.t. & \quad & c_{\tilde{j}} &= a_j
\\
 & \quad & \lambda_{j_1 \dots j_n} & \ge 0    
\end{align}
Clearly, this problem can be solved numerically by convex and quadratic optimization
solvers such as SDTP, Sedumi, or Yalmip. However, the number of coefficients $c_{i_1 \dots i_n}$ grows exponentially with the number of qubits. Therefore, a numerical approach
becomes practically intractable even for moderate qubit numbers. We will
therefore derive an analytical solution.

Without loss of generality, we restrict the measurement values
$a_k$ to be positive. Furthermore, we will only consider the case, in which
the measurement outcomes of neighboring qubits fulfil $a_{k_1}+a_{k_2} \ge 1$. 
In other words, we restrict to the case of a non-zero fidelity with the
desired stabilizer state.

Now one can easily check that the choice
\begin{align}
\label{Eq:coeff}
c_{i_1 \dots i_n} = \sum_k i_k a_k - \sum_k i_k + 1
\end{align} 
fulfils all the constraints in the above quadratic program. Positivity of
the eigenvalues follows from the fact that $\rho$ is diagonal in the stabilizer
basis. Thus, these coefficients represent a solution, which we denote by
$c^*$ in the following, to the primal problem.

Since the primal problem is convex, we may simply check the \textit{Karush-Kuhn-Tucker}
 (KKT) conditions for optimality of the solution\cite{14}. To be explicit, the KKT
conditions require
\begin{align}
\label{eq:KKT1}
\lambda_{j_1 ... j_n} (c^*) & \ge 0, 
\\
\label{eq:KKT2}
g(c_{\tilde{j}} = c_{\tilde{j}}^*) := c_{\tilde{j}} - a_j & = 0,
\\
\label{eq:KKT3} 
\mu_{j_1 \dots j_n} & \ge 0,
\\
\label{eq:KKT3}
\mu_{j_1 \dots j_n} \lambda_{j_1 ... j_n} (c^*) & = 0,
\\
\label{eq:KKT4}
\nabla_{j_1 \dots j_n} P(c^*) - \sum_{i_1, \dots, i_n=0}^1 \mu_{i_1 \dots i_n} \nabla_{j_1 \dots j_n} \lambda_{i_1 ... i_n} + \sum_{\tilde{m}=1}^n
\nu_{\tilde{m}} \nabla_{j_1 \dots j_n} g(c_{\tilde{m}}) & = 0.
\end{align}
The first two conditions state that $c^*$ is primal feasible, which
is already proved. The last condition guarantees the optimality of $c^*$.
Instead of  deriving the dual quadratic program, we will now prove that Lagrange
multipliers $\mu_{j_1 \dots j_n}$ and $\nu_{\tilde{m}}$ can always be found to the solution $c^*$, such that the KKT
conditions are fulfilled, thus showing that $c^*$ is indeed optimal.

A valid solution for the Lagrange multipliers is provided in the following
way: in order to satisfy (\ref{eq:KKT3}) choose $\mu_{j_1 \dots j_n} = 0$, if $\lambda_{j_1 \dots j_n}(c^*) > 0$.
 
Condition (\ref{eq:KKT4}) can be rewritten as
\begin{align}
\frac{2}{2^n}c - \frac{1}{2^n}A \mu + \nu = 0, 
\end{align} 
where it follows from Eq. \ref{eq:lambda} that $A$ is the normalized Hadamard matrix 
with elements $A_{i_1,\ldots,i_n,j_1,\ldots,j_n}=(-1)^{\sum_{k} i_kj_k}$, 
$c = (c_{j_1 \dots j_n})$, $\lambda = (\lambda_{j_1 \dots j_n})$, 
and $\nu$ is the vector containing $\nu_{\tilde{m}}$ for $\tilde{m} = 1,\dots,n$, while
all other entries $\nu_{j_1 \dots j_n}=0$. 
Here the matrix $A$ is chosen such that $A c
= 2^n\lambda$. Because of the orthogonality relation of the Hadamard matrix, 
and of $A = A^T$, it follows with $A^T A = 2^n I$:
\begin{align}
\mu = A^T (\frac{2}{2^n}c+\nu)= 2 \lambda+A \nu. 
\end{align}
If $\lambda_{\vec{j}} > 0$ ($\vec{j} = (j_1, \dots, j_n)$), then it must
hold that $\mu_{\vec{j}} = 0$. Especially, in any case $\mu_{\vec{0}} = 0$,
since the largest eigenvalue of $\rho$ is $\lambda_{\vec{0}}$. Therefore,
we have:
\begin{align}
\sum_{\vec{m} \in I} (-1)^{\vec{j} \cdot \vec{m}} \nu_{\vec{m}} = -2 \lambda_{\vec{j}}.
\end{align}
Here, $I$ denotes the set of bit-strings with at most one bit unequal zero,
as in the vector $\nu$ only the first $n$ entries are unequal zero.
More formally, $I = \{(0,\dots,0),(1,0,\dots,0),\dots,(0,\dots,0,1)\}$. 
This system of equations has the solution
\begin{align}
\nu_{\vec{l}} & = \lambda_{\vec{l}} - \lambda_{\vec{0}},~\vec{l} \neq \vec{0},
\\
\nu_{\vec{0}} & = -2 \lambda_{\vec{0}} - \sum_{\vec{l} \neq \vec{0}} \nu_{\vec{l}}.
\end{align}
It remains to show that $\mu_{\vec{j}} \ge 0$ for $\lambda_{\vec{j}}=0$:
\begin{align}
\mu_{\vec{j}} & = \sum_{\vec{m}} (-1)^{\vec{j} \cdot \vec{m}} \nu_{\vec{m}}
= \nu_0 + \sum_{\vec{m} \neq \vec{0}} (-1)^{\vec{j}\cdot \vec{m}} \nu_{\vec{m}}
\\
& = - 2 \lambda_{\vec{0}} - \sum_{\vec{l} \neq 0} \nu_l + \sum_{\vec{m} \neq
0} (-1)^{\vec{j}\cdot \vec{m}} \nu_{\vec{m}}
\\
& = - 2 \lambda_{\vec{0}} + \sum_{\vec{m} \neq
0} ((-1)^{\vec{j}\cdot \vec{m}}-1) \nu_{\vec{m}}
\\
& = - 2 \lambda_{\vec{0}}  + \sum_{\vec{m} \neq
0} ((-1)^{\vec{j}\cdot \vec{m}}-1) (\lambda_{\vec{m}}-\lambda_{\vec{0}})
\\
& = - 2 \lambda_{\vec{0}}  + \frac{1}{2^n} \sum_{\vec{m} \neq
0} ((-1)^{\vec{j}\cdot \vec{m}}-1) \sum_{\vec{i}} ((-1)^{\vec{i}\cdot \vec{m}}-1)
c_{\vec{i}}
\\
\label{eq:mupositivity}
& = - 2 \lambda_{\vec{0}} + \frac{1}{2^n} \sum_{\vec{m} \neq
0} (-2 \delta_{1,\vec{j}\cdot \vec{m}}) \sum_{\vec{i}} (-2 \delta_{1,\vec{i}\cdot \vec{m}}) c_{\vec{i}}.
\end{align}
Since it is clear that $\lambda_{\vec{0}} > 0$, we can restrict our attention
to eigenvalues $\lambda_{\vec{j}} = 0$ and dual variables $\mu_{\vec{j}}$  with at least one index $j_{\alpha}=1$. Therefore, we have
\begin{align}
\sum_{\vec{m} \neq 0} (-2 \delta_{1,\vec{j}\cdot \vec{m}}) \sum_{\vec{i}} (-2 \delta_{1,\vec{i}\cdot \vec{m}}) \frac{1}{2^n} c_{\vec{i}} \ge 4 \sum_{i_\beta=0,\beta
\in \{1,\dots,n\}-\alpha}^1 \frac{1}{2^n} c_{i_1 ... i_{\alpha-1} 1 i_{\alpha+1}...i_n}
\end{align}
The RHS of the last inequality can be rewritten as
\begin{align}
\frac{4}{2^n} \sum_{i_\beta=0,\beta
\in \{1,\dots,n\}-\alpha}^1  c_{i_1 ... i_{\alpha-1} 1 i_{\alpha+1}...i_n}
= 4 (\lambda_{\vec{0}} - \sum_{i_\beta=0,\beta
\in \{1,\dots,n\}-\alpha}^1 \frac{1}{2^n} c_{i_1 ... i_{\alpha-1} 0 i_{\alpha+1}...i_n}).
\end{align}
Employing Eq. (\ref{Eq:coeff}), the last term may be evaluated as
\begin{align}
\frac{1}{2^n} \sum_{i_\beta=0,\beta
\in \{1,\dots,n\}-\alpha}^1 c_{i_1 ... i_{\alpha-1} 0 i_{\alpha+1}...i_n}
& = \sum_{i_\beta=0,\beta
\in \{1,\dots,n\}-\alpha}^1 \frac{1}{2^n}(\sum_{k=1}^n i_k a_k - \sum_k i_k+1)
\\
& = \frac{1}{4} (\sum_{k=2}^n a_k-n+2) \le \frac{\lambda_{\vec{0}}}{2}.
\end{align}
The last inequality follows from the fact that $\lambda_{\vec{0}} = \frac{1}{2}
(\sum_{k=1}^n a_k-n+2)$. Applying this result to Eq. (\ref{eq:mupositivity})
immediately implies $\mu_{\vec{j}} \ge 0$, if $\lambda_{\vec{j}} = 0$. Therefore,
all KKT conditions are fulfilled, proving that the solution $c^*$ is indeed
optimal.
\section{Example: Two qubit purity}
For the purpose of illustration, let us consider the simple example of two
qubits, supposedly prepared as a cluster state. Given only the outcomes of
the stabilizer measurements $a_1
= tr(\rho X \otimes Z)$ and $a_2 = tr(\rho Z \otimes X)$, we seek the lowest
purity compatible with these two measurements. Using the symmetries of the
stabilizers, we may restrict the problem to density matrices of the form
\begin{align}
\rho = \frac{1}{4} (c_{00} \mathbf{1} + c_{10} K_1 + c_{01} K_2 + c_{11}
K_{1}K_2).
\end{align}
Because of $tr(\rho) = 1$, $c_{00} = 1$. Furthermore, the measurements determine
$c_{10} = a_1$ and $c_{01} = a_2$. In order to find the minimal purity, the
objective is now to minimize
\begin{align}
P_{min} = \frac{1}{4} (c_{00}^2 + c_{10}^2 + c_{01}^2 + c_{11}^2)
\end{align}
subject to $\lambda_{j_1 j_2}(c) = \sum_{i_1, i_2 = 0}^1 (-1)^{i_1 j_1+ i_2
j_2} c_{i_1 i_2} \ge 0$. 

As solution to the primal problem, one chooses $c_{11} = \frac{1}{4} (a_1+a_2-1)$.
Then the eigenvalues of $\rho$ are simply given by
\begin{align}
\lambda_{00} & = \frac{1}{2} (a_1+a_2)
\\
\lambda_{10} & = \frac{1}{2} (1-a_1)
\\
\lambda_{01} & = \frac{1}{2} (1-a_2)
\\
\lambda_{11} & = 0
\end{align}
thus fulfilling the positivity constraint. To see the optimality of the solution,
one can check the remaining KKT conditions.
\begin{align}
\label{eq:ExampleKKT1}
\frac{1}{2} & -\mu_1 - \mu_2 - \mu_3 - \mu_4 + \nu_1 & = 0,
\\
\frac{a_1}{2} & -\mu_1 + \mu_2 - \mu_3 + \mu_4 + \nu_2 & = 0, 
\\
\frac{a_2}{2} & -\mu_1 - \mu_2 + \mu_3 + \mu_4 + \nu_3 & = 0,
\\
\label{eq:ExampleKKT4}
\frac{1}{2} (a_1+a_2-1) & -\mu_1 + \mu_2 + \mu_3 - \mu_4  & = 0 
\end{align} 
The Lagrange multipliers $\nu$ are determined by
\begin{align}
\nu_{10} & = \lambda_{10}-\lambda_{00} = \frac{1}{2} (1-2a_1-a_2),
\\
\nu_{01} & = \lambda_{01}-\lambda_{00} = \frac{1}{2} (1-2a_2-a_1).
\\
\nu_{00} & = -2 \lambda_{00} - \nu_{10} - \nu_{01} = \frac{1}{2} (a_1+a_2)-1.
\end{align}
If the measurement outcomes fulfil $a_1, a_2 > 0$, which should be the case
in a cluster state experiment, then $\lambda_{00},\lambda_{10}, \lambda_{01}
> 0$, and $\lambda_{11} = 0$. Therefore, $\mu_1 = \mu_2 = \mu_3 = 0$, to
fulfill the KKT condition (\ref{eq:KKT3}). It is now straightforward to check
that the choice of these Lagrange multipliers satisfy conditions (\ref{eq:ExampleKKT1})
through (\ref{eq:ExampleKKT4}), thus proving optimality of the solution.

\section{Quality of the bounds}
\label{sec:Quality}
In this section we will test the usefulness of the obtained bound on the
purity. For this purpose, let us consider a perfect cluster state, which
is subject to dephasing for a certain time, which is here assumed to be 10
ms. The system is then described by a master equation of the form
\begin{align}
\label{eq:mastereq}
\dot{\rho} = \frac{\gamma}{2}\sum_i (Z_i \rho Z_i - \rho),
\end{align}
where $\gamma$ is the dephasing rate, which is assumed to be $(\unit[10]{ms})^{-1}$.

\begin{table}
\caption{Comparison between exact and estimated purities for noisy cluster
states}{\begin{tabular}{|c|c|c|c|}
\hline
No. qubits & exact purity & estimated purity & relative deviation \\
\hline
2 & 0.8269 & 0.8233 & 0.0044 \\
3 & 0.7520 & 0.7417 & 0.0137 \\
4 & 0.6838 & 0.6646 & 0.0281 \\
\hline
\end{tabular}}
\end{table}

\section{Estimation with error bars}
Needless to say, due to a finite number of measurements and experimental
imperfections, the measurement outcomes $a_i$ we considered in previous sections
posses an error $\Delta a_i \ge 0$, which clearly affects the result of a
purity estimation. In this section we briefly address
the question how to take into account such errors.  

Let us assume we measured the stabilizers $K_i$ with measurement outcomes
$a_i \pm \Delta a_i = tr(\rho K_i)$. As seen in Sec. \ref{sec:Lowest-Purity},
the optimal solution to the purity minimization problem (\ref{eq:Pmin}) is
given by a stabilizer diagonal representation of the density matrix with
coefficients $c_{i_1 ... i_n} = \sum_k i_k a_k - \sum_k i_k + 1$. From this,
it can be easily seen that the choice $c_{i_1 ... i_n}^{(\pm)} = \sum_k i_k (a_k \pm \Delta a_k) - \sum_k i_k + 1$ leads to upper (lower) error estimates
of the minimal purity with objective values $P_{min}^{(\pm)} = \frac{1}{2^n} \sum_{i_1,
..., i_n=0}^1 (c_{i_1... i_n}^{(\pm)})^2$.

\section{Entropy estimation}
\label{sec:entropy}
Alongside the purity of a quantum state $tr(\rho^2)$, the entropy is another quantifier for the degree of disorder of a system. Given only the measurement outcomes of the stabilizer, one can employ similar techniques as in the case of the purity to estimate the entropy.

The von Neumann entropy is given by $S=-tr(\rho log \rho)$. Again, let
us consider the generators of the stabilizer group as our observables with
measurement outcomes $a_1 = tr(\rho K_1), \dots, a_n = tr(\rho K_n)$. Then
we may restrict to states diagonal in the stabilizer basis $\rho = \frac{1}{2^n} \sum c_{i_1
\dots i_n} K_1^{i_1} \dots K_n^{i_n}$ with eigenvalues $\lambda_{j_1 \dots
j_n} (\rho) = \sum_{i_1, \dots, i_n = 0}^1 (-1)^{\sum_k i_k j_k} c_{i_1 \dots i_n}$. Now, we need to maximize the entropy over all states predicting the
measurement outcomes $a_k$, $k \in {1,...,n}$. 
\begin{align}
S_{max} = &max \{S(\rho) : tr(\rho K_i)=a_i, \rho \ge 0\}
\\
= & min \{ \sum_{i_1,\dots,i_n=0}^1 \lambda_{i_1 \dots i_n} log \lambda_{i_1 \dots i_n} : \lambda_{i_1 \dots i_n} \ge 0, c_{\tilde{j}} =a_j\}
\end{align}
Clearly, the solution $c_{i_1 \dots i_n} = \sum i_k a_k -\sum i_k+1$ satisfies
the constraints, thus providing a lower bound on the entropy. Considering
initially perfect cluster states subject to dephasing according to Eq. (\ref{eq:mastereq}),
one obtains the results given in the table \ref{tab:entropy}. Remarkably, the estimate differs only a few per cent from the exact value of the von Neumann entropy.

Even though the least purity state provides a very good lower bound on the
maximal entropy, an exact derivation of the entropy is possible in the following
way: given the probabilities $p_k^{(\pm)} = \frac{1 \pm a_k}{2}$, then it follows from the subadditivity
of the entropy $S(\{p_k\}) \le \sum_k S(p_k)$, that the maximal entropy is given by
\begin{align}
S_{max} = -\sum_{i_1, \dots, i_n=0}^1 \lambda_{i_1 \dots i_n} log \lambda_{i_1 \dots i_n},
\end{align}
with  $\lambda_{i_1 \dots i_n} = \prod_{k=1}^n \frac{1 + (-1)^{i_k} a_k}{2}$,
and the indices $a_k$ being a binary index.
\begin{table}
\caption{Comparison between exact and estimated entropies from the least purity
state for noisy cluster
states}{\begin{tabular}{|c|c|c|c|}
\hline
No. qubits & exact entropy & estimated entropy & relative deviation \\
\hline
2 & 0.3827 & 0.3803 & 0.0063 \\
3 & 0.5740 & 0.5667 & 0.0127 \\
4 & 0.7653 & 0.7505 & 0.0193 \\
\hline
\end{tabular}}
\label{tab:entropy}
\end{table}

\section{Conclusion}
In this paper we have developed techniques to estimate purities and entropies
in cluster state experiments with few measurements. We assume only that the
generators of the stabilizer group are measured in an experiment. Therefore,
the number of local measurements is linear in the qubit number, compared
to an exponential growth in the number of measurement settings for full-state
tomography. Given only these measurement outcomes, we have developed the
optimal analytical solution to the minimal purity compatible with the measurement
data. A comparison of this bound on the purity with the exact purity calculated
from simulated data of noisy cluster states shows that, despite the little number of measurements, the bound on the purity differs only by a few per
cent from the exact value. In a similar way, we obtained a lower bound on the von Neumann entropy considering only measurements of the stabilizers
of a cluster state. This bound deviates only by a few percent from the exact
value of the entropy, too. In addition, we employed the subadditivity of
the entropy to obtain an analytical result for the exact entropy of a stabilizer
state.   

\section*{Acknowledgments}
Harald Wunderlich thanks Miguel Navascues and Shashank Virmani for useful discussions. We acknowledge financial support from EU Integrated Project
QAP with contract number 015848, the Royal Society Wolfson Merit Award Scheme
and the EU STREP project HIP.

\vspace*{-6pt}   

\vspace*{-5pt}   

%
%
%
%
%
%
%
%
%
%



\end{document}